\PassOptionsToClass{natbib=false,balance=false}{acmart}
\documentclass[sigconf,screen,nonacm]{acmart}

\setcopyright{none}
\authorsaddresses{}
\makeatletter
\@ifundefined{hyxmp@parse@acmart}{}{\let\hyxmp@parse@acmart\relax}
\makeatother
\usepackage{graphicx}
\usepackage{microtype}
\usepackage{booktabs,tabularx}
\usepackage{amsmath}

\hypersetup{
  hidelinks,
  keeppdfinfo=true,
  pdfauthor={Juhyeong Jin, Hyunjun Cho, Joo-Young Kim},
  pdftitle={Synapse: An HBM-Coordinated HBF Near-Memory Computing Architecture for Pipelined Agentic AI Execution}
}

\graphicspath{{figures/}}

\makeatletter
\let\synapse@base@addtotoporbot\@addtotoporbot
\newcommand{\SynapseFloatColumn}[1]{%
  \expandafter\xdef\csname synapse@side@\number\@currbox\endcsname{#1}%
}
\def\@addtotoporbot{%
  \@ifundefined{synapse@side@\number\@currbox}{%
    \synapse@base@addtotoporbot
  }{%
    \edef\synapse@side{\csname synapse@side@\number\@currbox\endcsname}%
    \def\synapse@right{R}%
    \ifx\synapse@side\synapse@right
      \if@firstcolumn\else\synapse@base@addtotoporbot\fi
    \else
      \if@firstcolumn\synapse@base@addtotoporbot\fi
    \fi
    \if@insert
      \expandafter\global\expandafter\let\csname synapse@side@\number\@currbox\endcsname\relax
    \fi
  }%
}
\makeatother
\makeatletter
\def\@subsecfont{\bfseries\normalsize\section@raggedright}
\makeatother
\DeclareCaptionFont{synapsecaption}{\fontsize{8}{9.5}\selectfont}
\newcommand{\para}[1]{\noindent\textbf{#1}\ }

\renewcommand{\topfraction}{0.90}

\renewcommand{\textfraction}{0.08}
\renewcommand{\floatpagefraction}{0.80}

\makeatletter
\providecommand{\@IEEEtablecaptionsepspace}{\vskip\abovecaptionskip\relax}
\providecommand{\@IEEEtabletopskipstrut}{\ifdim\prevdepth=-1000pt\rule{0pt}{\topskip}\fi}
\makeatother

\providecommand{\BIBdecl}{}
\makeatletter
\g@addto@macro\BIBdecl{%
  \microtypesetup{protrusion=false}%
  \ifdefined\UrlBreaks
    \expandafter\def\expandafter\UrlBreaks\expandafter{\UrlBreaks
      \do\a\do\b\do\c\do\d\do\e\do\f\do\g\do\h\do\i\do\j
      \do\k\do\l\do\m\do\n\do\o\do\p\do\q\do\r\do\s\do\t
      \do\u\do\v\do\w\do\x\do\y\do\z
      \do\A\do\B\do\C\do\D\do\E\do\F\do\G\do\H\do\I\do\J
      \do\K\do\L\do\M\do\N\do\O\do\P\do\Q\do\R\do\S\do\T
      \do\U\do\V\do\W\do\X\do\Y\do\Z
      \do\0\do\1\do\2\do\3\do\4\do\5\do\6\do\7\do\8\do\9
      \do\-}%
    \Urlmuskip=0mu plus 1mu\relax
  \fi
  \emergencystretch=1em\relax
}
\makeatother

\title{Synapse: An HBM-Coordinated HBF Near-Memory Computing Architecture for Pipelined Agentic AI Execution}
\author{Juhyeong Jin}
\email{pearlbro@kaist.ac.kr}

\author{Hyunjun Cho}
\email{h.cho@kaist.ac.kr}

\author{Joo-Young Kim}
\email{jooyoung1203@kaist.ac.kr}

\renewcommand{\shortauthors}{Juhyeong Jin, Hyunjun Cho, and Joo-Young Kim}

\makeatletter
\newcommand{\synapse@authorblock}[2]{%
  \begin{minipage}[t]{0.30\textwidth}
    \centering\normalfont
    {\fontsize{11}{13}\selectfont #1\par}%
    \vspace{2pt}%
    {\fontsize{10}{12}\selectfont KAIST\par}%
    \vspace{1pt}%
    {\fontsize{9}{11}\selectfont #2\par}%
  \end{minipage}%
}
\def\@mkauthors@iii{%
  \global\setbox\mktitle@bx=\vbox{%
    \unvbox\mktitle@bx\par\medskip
    \noindent\hfill
    \synapse@authorblock{Juhyeong Jin}{pearlbro@kaist.ac.kr}\hfill
    \synapse@authorblock{Hyunjun Cho}{h.cho@kaist.ac.kr}\hfill
    \synapse@authorblock{Joo-Young Kim}{jooyoung1203@kaist.ac.kr}\hfill\null
    \par\medskip
  }%
}
\makeatother

\begin{document}

\raggedbottom

\begin{abstract}
\fontsize{8}{10}\selectfont
During multi-agent video reasoning, model weights and per-agent key--value caches can exceed high-bandwidth memory (HBM) capacity.
High-bandwidth flash (HBF) adds capacity, but agents still wait for GPU execution.
\textbf{Synapse}, an HBM-coordinated near-memory computing architecture, overlaps HBF attention and expert execution with GPU computation across agents.
It also reduces dispatch delays and unnecessary refreshes.
Across three models, \textbf{Synapse} achieves 48.92--66.60\% lower end-to-end latency, up to $2.98\times$ throughput, and up to 64.48\% lower modeled energy per output token than GPU execution with passive HBF.
Wear simulations project 46\% longer HBF lifetime than block-wide refresh at equal workload rates.
\end{abstract}

\maketitle
\hypersetup{pdfcreator={LaTeX with acmart and hyperref}}
\pagestyle{plain}

\ifdefined\SynapseWorkflowFigureQueued\expandafter \fi
\global\let\SynapseWorkflowFigureQueued\relax
\begin{figure}[!t]
\SynapseFloatColumn{R}
\centering
\includegraphics[width=\columnwidth]{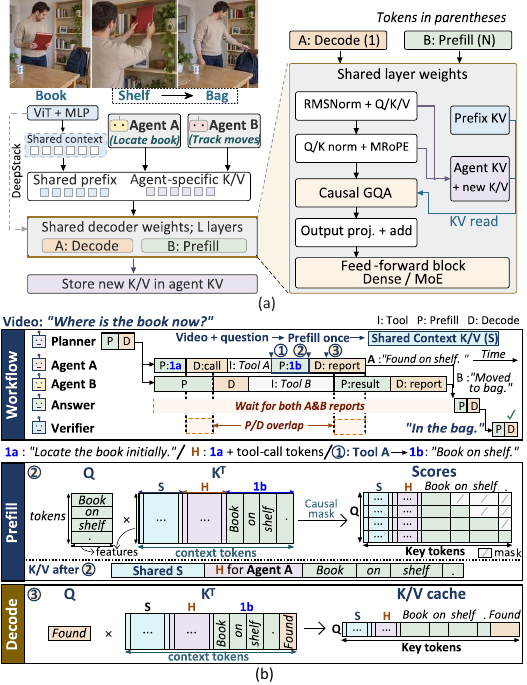}
\caption{Shared-context multi-agent video reasoning:
(a) shared weights and prefix KV, agent-specific KV, and decoder-layer operations;
(b) tool calls and agent dependencies, with overlapping prefill/decode and KV reuse.}
\Description{Shared-context multi-agent video reasoning:
(a) shared weights and prefix KV, agent-specific KV, and decoder-layer operations;
(b) tool calls and agent dependencies, with overlapping prefill/decode and KV reuse.}
\label{fig:synapse_workflow}
\end{figure}

\section{Introduction}\label{sec:intro}

Video reasoning is emerging as a key capability of AI assistants, enabling them to answer questions about events and changes over time.
Google's agentic video understanding~\cite{gemini_agentic_video2026} interleaves model reasoning with video-tool calls.
These calls retrieve relevant segments, allowing the model to connect evidence across time.
Qwen2.5-VL~\cite{qwen25vl_report} and Qwen3-VL~\cite{qwen3vl_report} report higher video-understanding and reasoning accuracy for their larger variants.
These results motivate using large models for video reasoning.
Mixture-of-experts (MoE) architectures make this scale practical by activating only a few experts per token.

Video reasoning is also increasingly agentic: systems such as A4VL~\cite{a4vl2026} divide a question across multiple agents that share a common model. However, serving such a model to multiple agents exceeds high-bandwidth memory (HBM) capacity.
A Qwen3-VL 235B model's BF16 weights alone require about 470\,GB, about $5.9\times$ the 80\,GB HBM capacity, before accounting for key--value (KV) caches that each agent retains through tool pauses.
Although sharding the model across multiple GPUs can aggregate their HBM capacity, six 80\,GB GPUs provide only 480\,GB of raw capacity, leaving little room for per-agent KV caches, activations, and communication buffers.
Using additional GPUs alleviates this capacity pressure but introduces communication and synchronization overhead from tensor or expert parallel execution.
This trade-off motivates a high-capacity memory tier to complement HBM.
High-Bandwidth Flash (HBF), specified by OCP~\cite{ocp}, provides additional capacity with high read bandwidth.

HBF can alleviate this capacity pressure, but using it only as a high-bandwidth backing store does not ensure timely agent progress.
Fig.~\ref{fig:synapse_workflow} illustrates our target workflow for ``\textit{Where is the book now?}'', an object-location question found in Ego4D’s episodic-memory benchmark~\cite{ego4d_cvpr2022}.
Agent A locates the book and Agent B tracks its movement through separate calls to the same model.
They share model weights and a context prefix but retain separate conversation KV through tool pauses (Fig.~\ref{fig:synapse_workflow}(a)).
Answer combines their reports, and Verifier checks the evidence (Fig.~\ref{fig:synapse_workflow}(b)).
When B's tool result arrives, B prefills it while A decodes its report.
With passive HBF, both agents rely on the GPU for attention and expert computation.
This execution pattern exposes a central challenge: HBF expands capacity, but agents still contend for the GPU. 
The goal is to enable low-latency, reliable execution of multi-agent video reasoning.

Achieving this goal requires addressing four system-level concerns.
First, GPU contention stalls A’s attention when B’s prefill occupies the GPU, even though A’s active KV is already resident in HBM.
Second, A’s expert computation encounters two consecutive bottlenecks: weight-read requests from B and other agents delay the delivery of A’s weights from HBF, while B’s prefill and other GPU work delay A’s expert execution even after the weights reach HBM.
Third, when HBF execution relies on GPU commands and software readiness updates, GPU work can delay HBF issue, and software updates can delay dependent operations even after results reach HBM.
Fourth, as agents resume after tool calls and reread shared weights, NAND read disturbance accumulates. Refreshing the affected weights adds NAND reads and writes. These additional refresh writes gradually wear out NAND flash cells, shortening HBF lifespan.

\ifdefined\SynapseWorkflowFigureQueued\expandafter \fi
\global\let\SynapseWorkflowFigureQueued\relax
\begin{figure}[!t]
\SynapseFloatColumn{R}
\centering
\includegraphics[width=\columnwidth]{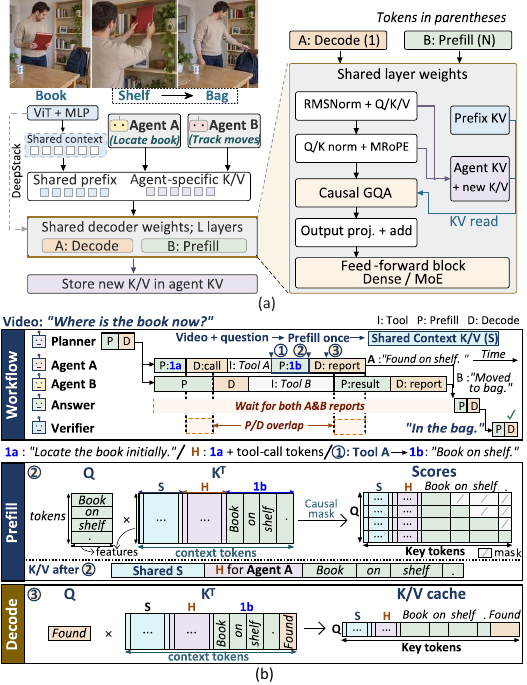}
\caption{Shared-context multi-agent video reasoning:
(a) shared weights and prefix KV, agent-specific KV, and decoder-layer operations;
(b) tool calls and agent dependencies, with overlapping prefill/decode and KV reuse.}
\Description{Shared-context multi-agent video reasoning:
(a) shared weights and prefix KV, agent-specific KV, and decoder-layer operations;
(b) tool calls and agent dependencies, with overlapping prefill/decode and KV reuse.}
\label{fig:synapse_workflow}
\end{figure}

\begin{figure}[!t]
\SynapseFloatColumn{R}
  \centering
  \includegraphics[width=\columnwidth]{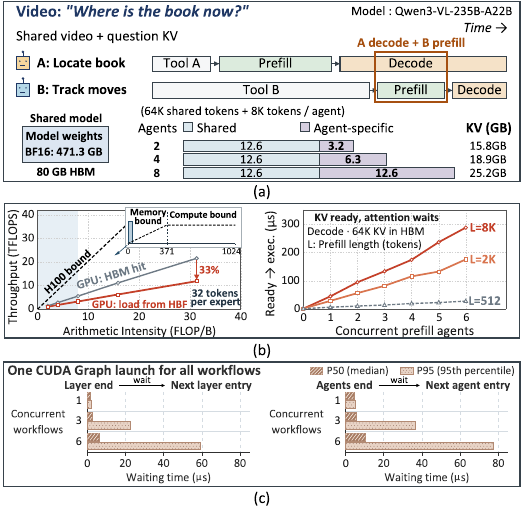}
   \caption{Prior-work limits from capacity to scheduling: (a) model/KV capacity;
(b) modeled weight loading and attention waits with HBM-ready KV;
(c) layer/agent transition gaps (P50, P95) with one CUDA Graph launch.}
\Description{Prior-work limits from capacity to scheduling: (a) model/KV capacity;
(b) modeled weight loading and attention waits with HBM-ready KV;
(c) layer/agent transition gaps (P50, P95) with one CUDA Graph launch.}
  \label{fig:synapse_motivation}
\end{figure}

We present \textbf{Synapse}, an HBM-coordinated HBF near-memory computing architecture that advances A’s decode on HBF while the GPU runs B’s prefill. 
Synapse consists of three components. 
A coordinator on the HBM base die comprises an Operand Issue Engine (OIE), which launches ready HBF work without waiting for the GPU, and a Result Writeback Engine (RWE), which updates successor readiness once results become visible in HBM, without software intervention. 
A compute engine on the HBF base die comprises a Shared MAC Engine (SME) and a Vector Function Engine (VFE), which serve both attention and expert computation within the base-die budget. 
An HBF-side protection engine uses counters, comparators, and a sequencer to enforce wordline stress limits and refresh shared weights locally. 
Together, these components enable concurrent agent progress in four ways:

\begin{itemize}
\item \textit{\textbf{Flow}} executes decode attention on HBF using HBM-resident operands, allowing one agent’s decode to progress while the GPU prefills another agent.
\item \textit{\textbf{Fuse}} executes MoE experts beside NAND-resident weights and shares one weight stream across calls to the same expert, avoiding expert-weight transfers to the GPU and overlapping expert computation with GPU work.
\item \textit{\textbf{Forward}} reduces issue and completion delays: OIE launches ready work without waiting for a GPU command, while RWE marks the next dependent task ready as soon as result writeback completes.
\item \textit{\textbf{Fortify}} protects HBF lifetime by tracking actual physical reads per wordline and triggering refresh only when safety limits require it. Shared reads are counted once, and required refresh overlaps \textit{Flow} to hide maintenance latency.
\end{itemize}

Compared with GPU execution with passive HBF across three models, Synapse reduces end-to-end latency by 48.92--66.60\% and delivers up to $2.98\times$ the throughput.
Modeled energy per output token falls by up to 64.48\%.
At the same workload rate, wear analysis projects about 46\% longer lifespan over a baseline that triggers a block-wide refresh when any wordline reaches its stress limit.

\ifdefined\SynapseArchitectureFigureQueued\expandafter \fi
\global\let\SynapseArchitectureFigureQueued\relax
\setlength{\textfloatsep}{7pt plus 1pt minus 1pt}
\setlength{\dbltextfloatsep}{4pt plus 1pt minus 1pt}
\setlength{\floatsep}{5pt plus 1pt minus 1pt}
\renewcommand{\topfraction}{.95}
\renewcommand{\dbltopfraction}{.95}
\renewcommand{\textfraction}{.05}
\setcounter{topnumber}{2}
\renewcommand{\floatpagefraction}{.80}
\makeatletter
\setlength{\@fptop}{0pt}
\setlength{\@fpsep}{5pt}
\setlength{\@fpbot}{0pt plus 1fil}
\makeatother

\begin{figure*}[!t]
\centering
\includegraphics[width=\textwidth]{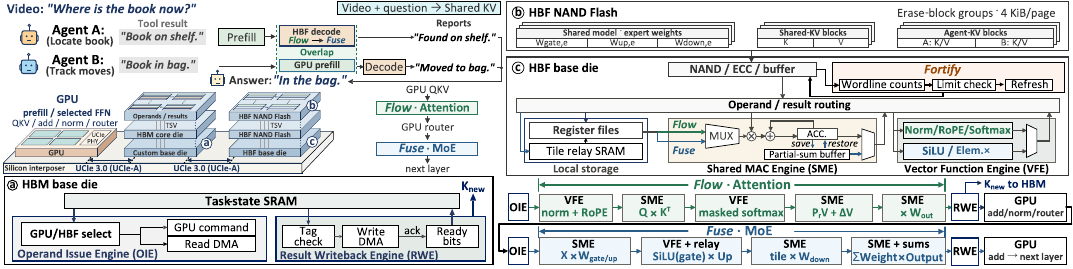}
\caption{Synapse architecture (one HBM--HBF pair): (a) HBM coordinator, (b) NAND storage, and (c) shared HBF compute engines.}
\Description{Synapse architecture (one HBM--HBF pair): (a) HBM coordinator, (b) NAND storage, and (c) shared HBF compute engines.}
\label{fig:synapse_architecture}
\end{figure*}

\section{Motivation and Related Work}
\label{sec:motivation}

\subsection{Memory Capacity Alone Does Not Ensure Concurrent Execution}
\label{sec:motivation_capacity}
\label{sec:motivation_state}

Fig.~\ref{fig:synapse_motivation}(a) shows agents sharing a model and context, overlapping prefill and decode, and retaining KV through tool pauses.
Qwen3-VL-235B-A22B-Instruct weights exceed 80\,GB HBM capacity; agent-specific KV grows with concurrency.
Duplex~\cite{yun2024duplex} combines an xPU with HBM-based processing-in-memory (PIM) for MoE experts and attention.
With CPU/SSD or passive HBF backing, nonresident expert weights still require loading into HBM; expert partial sums return to the xPU to be summed.
ReKV~\cite{rekv} retrieves query-relevant KV blocks; LiveVLM~\cite{livevlm} compresses streaming KV before retrieval.
Hot--Cold~\cite{hotcold2026} keeps active KV in HBM and paused KV in HBF.
These KV-management methods reduce HBM demand but still rely on the GPU for attention.

The left plot in Fig.~\ref{fig:synapse_motivation}(b) compares H100 expert execution with HBM-resident weights or modeled HBF loading.
More tokens per expert increase weight reuse and arithmetic intensity.
Decode adds one token per agent per step; longer histories enlarge KV, and routing can split tokens across experts.
Thus, experts often process batches of eight or fewer tokens (shaded region), limiting weight reuse and H100 compute utilization~\cite{synapse_h100_arch}.
The red curve includes the time to load 36~MiB of weights from HBF to HBM for each expert execution~\cite{ocp}.
Even at 32 tokens per expert, throughput is 33\% lower than with weights already in HBM.
Pre-gated MoE~\cite{pregated} overlaps next-layer weight transfer with computation but still delivers weights to the GPU.
This motivates computing experts near HBF-resident weights and reducing intermediate data sent to the GPU.

Beyond weight delivery, attention can wait for GPU execution even
when its KV is already in HBM.\
The right plot in Fig.~\ref{fig:synapse_motivation}(b) shows how long single-token decode attention waits to start on the GPU, even when its 64K-token KV cache is already in HBM.
With six concurrent prefills, decode attention waits an average of 28, 176, and 288~\textnormal{\textmu s} to start on the GPU for prefill lengths $L=512$, $2\mathrm{K}$, and $8\mathrm{K}$ tokens, respectively.
GPU markers show that attention waits behind prefill even when its KV is already in HBM.
This motivates running attention on HBF using KV in HBM while the GPU runs another agent's prefill.

\subsection{Data Readiness Does Not Guarantee Timely Progress}
\label{sec:motivation_delivery}
\label{sec:motivation_control}

Fig.~\ref{fig:synapse_motivation}(c) shows delays between dependent operations on H100 with $C=1,3,6$ concurrent workflows, measured over 15 runs per setting.
Across concurrency levels, inputs remain in HBM, and a single launch of a pre-captured CUDA Graph runs all workflows while preserving layer and agent dependencies.
GPU timestamps measure how long the next layer or agent waits after all required work finishes.
Intervals are pooled across workflows and runs for P50/P95.
At $C=6$, layer/agent gaps reach 6/11~\textnormal{\textmu s} at P50 and approximately 59/77~\textnormal{\textmu s} at P95.
These gaps include the overhead of CUDA kernels recording boundary timestamps.
Their growth with concurrency shows that HBM-resident inputs and a single graph launch still leave dependency gaps that can extend the critical path.

These gaps motivate coordinating successor readiness and HBF issue alongside GPU batching.
Sarathi-Serve~\cite{sarathi_osdi2024}, a software scheduler, batches chunked prefills with decodes, but does not coordinate HBM writeback with independent HBF issue.
Under GPU-mediated control, ready HBF work can wait behind another agent's prefill.
This motivates HBM-side control that marks successors ready when all required results are visible and issues HBF work without another GPU command.

\ifdefined\SynapseMotivationPreview\else

\fi

\subsection{Repeated Weight Reads Increase Flash Wear}
\label{sec:motivation_endurance}

Expert execution in HBF repeatedly reads NAND, causing read disturbance and eventual refresh writes.
Concurrent agents can share weight reads; independently progressing agents may reread the same weights.
STRAW~\cite{fortify_straw} reduces unnecessary reclaim through wordline-stress tracking; FLINT~\cite{fortify_flint} isolates refresh using an additional plane.
Neither coordinates maintenance with cross-agent expert-weight reuse or execution.
This motivates expert-aware maintenance that counts shared physical reads once and schedules refresh during attention, keeping NAND available for experts while meeting safety deadlines.
\ifdefined\SynapseFlowFuseFigureQueued\expandafter \fi
\global\let\SynapseFlowFuseFigureQueued\relax
\begin{figure}[!t]
\SynapseFloatColumn{R}
\centering
\includegraphics[width=\columnwidth]{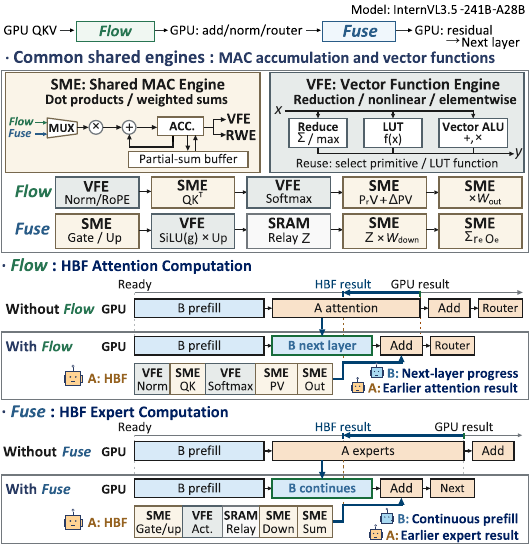}
\caption{Shared HBF engines for \emph{Flow} and \emph{Fuse}, allowing agent A to compute while agent B uses the GPU.}
\Description{Shared HBF engines for \emph{Flow} and \emph{Fuse}, allowing agent A to compute while agent B uses the GPU.}
\label{fig:synapse_flow_fuse}
\label{fig:synapse_flow}
\label{fig:synapse_fuse}
\end{figure}

\section{Synapse Architecture}
\label{sec:architecture}

Synapse advances one agent on HBF while another uses the GPU.
\emph{Flow} and \emph{Fuse} share HBF engines for attention and MoE expert computation; \emph{Forward} coordinates operand issue, writeback, and successor readiness; \emph{Fortify} protects shared weights against read disturb.

\subsection{Organization and Shared Engines}
\label{sec:arch_overview}
\label{sec:synapse_organization}

\noindent\textbf{Organization and data placement.}
Fig.~\ref{fig:synapse_architecture} separates the HBM coordinator (a), NAND storage (b), and HBF datapath (c).
The GPU executes prefill, QKV projection, routing, and residual addition.
HBM holds active weights, KV, operands, and results; HBF provides backing storage for weights and retained KV~\cite{ocp}.
Weights, shared-prefix KV, and agent-specific KV occupy separate erase-block groups.
The HBM base die hosts the Operand Issue Engine (OIE), Result Writeback Engine (RWE), and task state.
The HBF base die provides the Shared MAC Engine (SME), Vector Function Engine (VFE), and intermediate storage.

\noindent\textbf{End-to-end layer flow.}
In Fig.~\ref{fig:synapse_flow_fuse}, the GPU first computes QKV, and \emph{Flow} then performs attention on HBF.
The GPU applies residual addition and normalization, routes tokens to \emph{Fuse} for expert computation on HBF, and adds the expert outputs to the residual.
OIE supplies operands and RWE writes results to HBM, preserving each agent's dependencies while different agents progress concurrently on the GPU and HBF.

\subsection{\emph{\textbf{Flow}} and \emph{\textbf{Fuse}}---Shared HBF Execution}
\label{sec:arch_flow_fuse}
\label{sec:arch_flow}
\label{sec:arch_relay}
\label{sec:arch_fuse}
\label{sec:arch_reuse}

\ifdefined\SynapseFlowFuseFigureQueued\expandafter \fi
\global\let\SynapseFlowFuseFigureQueued\relax
\begin{figure}[!t]
\SynapseFloatColumn{R}
\centering
\includegraphics[width=\columnwidth]{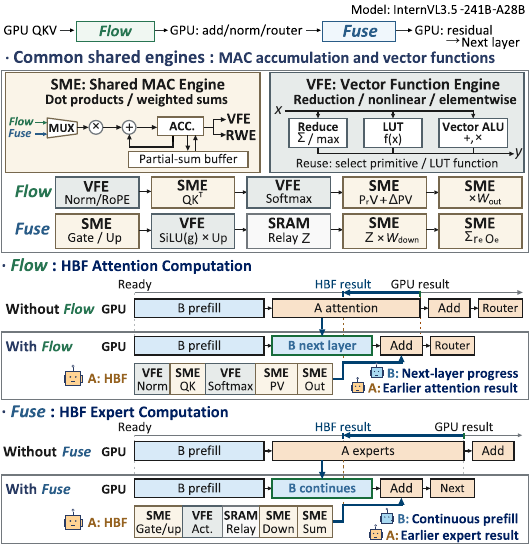}
\caption{Shared HBF engines for \emph{Flow} and \emph{Fuse}, allowing agent A to compute while agent B uses the GPU.}
\Description{Shared HBF engines for \emph{Flow} and \emph{Fuse}, allowing agent A to compute while agent B uses the GPU.}
\label{fig:synapse_flow_fuse}
\label{fig:synapse_flow}
\label{fig:synapse_fuse}
\end{figure}

\noindent\textbf{Shared compute engines.}
The Shared MAC Engine (SME) and Vector Function Engine (VFE) support both \emph{Flow} and \emph{Fuse}, avoiding separate attention and expert compute arrays.
SME reuses one multiply--accumulate array for attention products, expert projections, and weighted sums.
An operand multiplexer selects inputs, while accumulators retain partial results across tiles.
VFE provides reductions, function lookup tables, and elementwise arithmetic for normalization, RoPE, softmax, and SiLU gating.
Register files exchange intermediate values; local SRAM buffers attention scores, activated tiles, and partial sums.
Stage control selects operands, vector functions, and accumulation destinations.
Fig.~\ref{fig:synapse_flow_fuse} shows A's HBF computation overlapping B's GPU prefill.
\emph{Forward} handles operand issue and result writeback; dependent GPU work proceeds once its inputs are ready and the GPU is available.

\noindent\textbf{\emph{Flow}: attention through shared engines.}
OIE supplies QKV from HBM.
VFE normalizes Q and K and applies RoPE; SME computes attention scores, which VFE converts to probabilities through masked softmax.
SME then computes $P_{\rm r}V+\Delta P\,V$, where $P_{\rm r}$ denotes the rounded attention probabilities and $\Delta P=P-P_{\rm r}$ is the rounding residual.
Both products reuse the same buffered V tile, avoiding a second HBM fetch.
Alternating V buffers overlap transfers with computation.
SME performs the output projection using HBM-resident weights, and RWE writes the attention result to HBM.

\noindent\textbf{\emph{Fuse}: expert computation through shared engines.}
After GPU routing, OIE supplies hidden states and expert assignments from HBM.
Compatible ready inputs targeting the same model, layer, and expert share a NAND weight stream.
SME computes gate and up projections; VFE produces $Z=\operatorname{SiLU}(g)\odot u$.
Relay SRAM distributes each activated tile to channel registers holding different down-projection columns.
SME computes the down projections and accumulates expert outputs weighted by router scores.
RWE writes tagged results to HBM for residual addition on the GPU.
Expert weights and intermediate activations remain in HBF.
Both paths exchange inputs and results with the GPU through HBM, where \textit{Forward} coordinates task issue and updates successor readiness as results become visible.

\subsection{\emph{\textbf{Forward}}---HBM-Local Issue and Completion}
\label{sec:arch_forward}

\begin{figure}[!t]
\SynapseFloatColumn{L}
\centering
\includegraphics[width=\columnwidth]{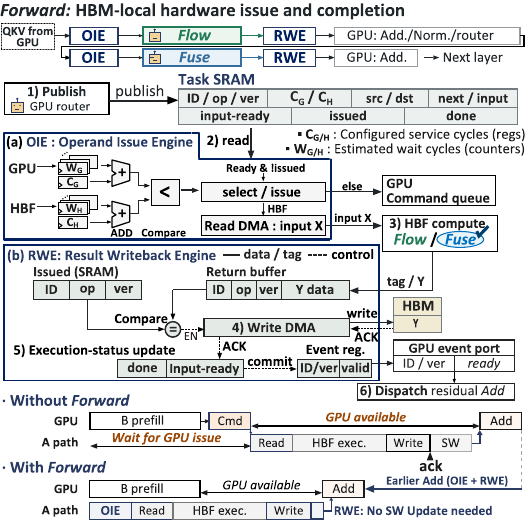}
\caption{\textit{Forward}: HBM-side task issue and completion.
(a) OIE selects GPU or HBF execution for ready tasks.
(b) RWE validates result tags and updates successor readiness after HBM writeback.}
\Description{\textit{Forward}: HBM-side task issue and completion.
(a) OIE selects GPU or HBF execution for ready tasks.
(b) RWE validates result tags and updates successor readiness after HBM writeback.}
\label{fig:synapse_forward}
\label{fig:synapse_execution}
\label{fig:synapse_agent_execution}
\end{figure}

\noindent\textbf{Publish and select (1--2).}
In Fig.~\ref{fig:synapse_forward}, the GPU publishes A's \emph{Fuse} descriptor to Task SRAM before B's prefill, after $X(A)$ and descriptor fields become visible (1).
Its dependent Add waits in the GPU queue for inputs.
Each descriptor stores operation metadata, data and dependency mappings, execution state, and $C_G,C_H$.
$W_G$ and $W_H$ estimate when the GPU and HBF will be available; $C_G$ and $C_H$ estimate the corresponding task execution times.
OIE sends a ready task awaiting dispatch to HBF if $W_H+C_H < W_G+C_G$; otherwise, it sends it to the GPU (2).

\noindent\textbf{Issue and execute (2--3).}
Without \emph{Forward}, agent A's HBF request waits until agent B's GPU prefill finishes.
With \emph{Forward}, OIE starts Read DMA for $X(A)$ while B's prefill continues on the GPU (2).
Once the command is accepted, OIE sets the entry's issued bit to prevent reissue.
HBF then runs \emph{Flow} or \emph{Fuse} on SME and VFE (3) and returns the task tag and $Y(A)$ to the RWE return buffer.
The comparison keeps placement, computation, and transfer times fixed, isolating the benefit of earlier issue.

\noindent\textbf{Write back and resume (4--6).}
RWE validates the buffered tag against the issued entry.
On a match, Write DMA writes $Y(A)$ to HBM and acknowledges completion only after $Y(A)$ becomes visible to GPU loads (4).
Execution-Status marks the \emph{Fuse} task as done and the $Y$ dependency of \emph{Add} as ready in Task SRAM (5).
Once $Y(A)$ is visible, \emph{Forward} holds its producer ID/version until the GPU queue accepts it and marks the corresponding dependency of \emph{Add} ready, without software polling or resubmission.
Once all inputs are ready and the GPU is available, the queue dispatches \emph{Add} if not already issued (6).
\emph{Add} computes $\mathrm{residual}+Y(A)$ without OIE reselection or software intervention.
The same path supports \emph{Flow} and \emph{Fuse} across layers.

\subsection{\emph{\textbf{Fortify}}---Hardware Read-Disturb Protection}
\label{sec:arch_fortify}

\begin{figure}[!t]
\SynapseFloatColumn{R}
\centering
\includegraphics[width=\columnwidth]{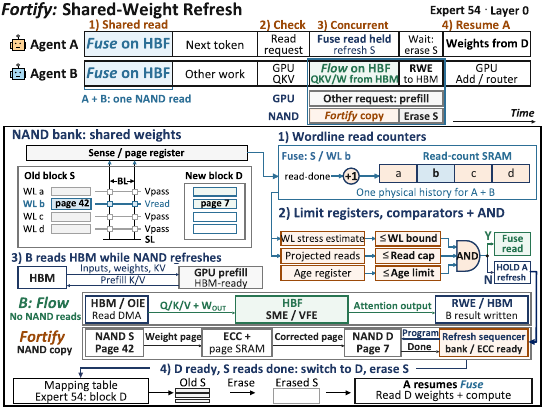}
\caption{\emph{Fortify}: per-wordline physical-read tracking and local refresh of shared weights, overlapped with \emph{Flow}.}
\Description{\emph{Fortify}: per-wordline physical-read tracking and local refresh of shared weights, overlapped with \emph{Flow}.}
\label{fig:synapse_fortify}
\end{figure}

\noindent\textbf{\emph{Fuse}-coupled physical-read accounting.}
\emph{Fortify} monitors physical NAND reads at each HBF bank rather than counting logical requests from individual agents.
In Fig.~\ref{fig:synapse_fortify}, \emph{Fuse} serves both agents with a single physical page read, so the bank produces one read-completion event.
The event identifies the selected wordline and increments its entry in the read-count SRAM.
A later token or resumed agent that physically rereads the weights produces another increment.
Hardware then estimates the read-disturb exposure imposed on the unselected wordlines by these physical reads.

\noindent\textbf{Comparator-gated NAND access.}
Before issuing the next NAND read, \emph{Fortify} projects its effect on wordline stress and physical-read count.
Parallel comparators check the projected values and retention age against hardware limits, and their combined result directly gates the NAND read-enable signal.
A safe request enters the NAND array, whereas a failed check holds the tagged \emph{Fuse} request and sets the bank's refresh-pending latch.
Thus, the protection decision is enforced in the NAND command path without software intervention or agent-level scheduling.

\noindent\textbf{\emph{Flow}-overlapped local refresh.}
The pending latch activates the refresh sequencer when the NAND bank, ECC engine, and page SRAM are available.
The sequencer reads each valid page from block S, corrects it through ECC, and programs the corrected page into block D.
Meanwhile, B's \textit{Flow} performs attention and output projection on the SME/VFE using Q/K/V and $W_{\mathrm{out}}$ from HBM, without accessing NAND.
Once the copy completes and outstanding reads drain, the mapping for Expert 54 is updated to D, and S is erased.
A's held \textit{Fuse} request then resumes using the refreshed weights in D.
This allows NAND refresh to overlap with B's computation while keeping read-disturb protection in hardware.
\section{Evaluation}
\label{sec:eval}

\subsection{Methodology}
\begin{table}[t]
\SynapseFloatColumn{L}
\makeatletter
\long\def\@makecaption#1#2{%
  \footnotesize\bgroup\par\centering
  \@IEEEtabletopskipstrut
  {\normalfont\footnotesize #1:\nobreakspace\scshape #2}\par
  \addvspace{0.5\baselineskip}\egroup
  \@IEEEtablecaptionsepspace
}
\makeatother
\centering
\caption{Experimental setup}
\label{tab:eval_setup}
\small
\setlength{\tabcolsep}{3pt}
\begin{tabular}{@{}p{0.23\columnwidth}p{\dimexpr0.77\columnwidth-6pt\relax}@{}}
\hline
\textbf{Component} & \textbf{Configuration} \\
\hline
HBM & $5\times16$\,GB HBM2e stacks (80\,GB total) \\
HBF & $5\times512$\,GiB stacks (2.5\,TiB total) \\
CPU memory & 512\,GB DRAM \\
CPU--GPU & PCIe Gen5 $\times16$; H2D/D2H: 55.7/46.1\,GB/s \\
D2D bandwidth & Max 1.536\,TB/s (UCIe, OCP Grade 2) \\
HBF compute & 2,048 BF16 MACs total at 400\,MHz \\
On-chip storage & 32\,MiB data path, incl.\ NAND page buffers \\
NAND flash & \raggedright
2,048 active banks; 4\,KiB pages\newline
$t_R=3\,\mu\mathrm{s}$; $t_{\mathrm{PROG}}=100\,\mu\mathrm{s}$
\tabularnewline
\hline
Access energy & NAND read/program: 6.8/100\,pJ/bit;
D2D: 6.8\,pJ/bit \\
Power (W) & Host: 100; HBF total background/active logic: 5/16 \\
\hline
\end{tabular}
\end{table}

\noindent\textbf{Experimental setup.}
GPU execution times are measured on an NVIDIA H100; HBF computation, NAND accesses, transfers, and contention are simulated using the parameters in Table~\ref{tab:eval_setup}.
Prior-work comparisons use Qwen3-VL-235B-A22B-Instruct~\cite{qwen3vl235_config}; ablations also include InternVL3.5-241B-A28B~\cite{internvl35_241_config} and GLM-5.3-Flash~\cite{glm53_flash_config}.

\begin{figure}[!t]
\SynapseFloatColumn{L}
\centering
\includegraphics[width=\columnwidth]{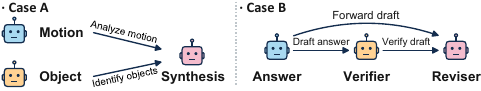}
\caption{Agent workflows.
In Case A, parallel motion and object analyses are combined into a report.
In Case B, an answer is drafted, verified against the video, and revised using both the draft and verification result.}
\Description{Agent workflows.
In Case A, parallel motion and object analyses are combined into a report.
In Case B, an answer is drafted, verified against the video, and revised using both the draft and verification result.}
\label{fig:eval_cases}
\end{figure}

\noindent\textbf{Workloads and baselines.}
Fig.~\ref{fig:eval_cases} shows the two evaluated workflows; each agent retains a separate KV cache.
Each prior-work comparison run replays Qwen3-VL traces from three LongVideoBench~\cite{longvideobench} validation videos, with four replays per video and 12 workflow instances in total.
They use 256 frames per video; ablation studies use 64.
The first two agents each output up to 64 tokens; the final agent, 32.
Workflows arrive every 10\,ms.
Each run processes all 12 instances, with at most $C$ active concurrently; $C=1$ executes workflows sequentially.
The default is $C=6$, and sweeps use $C\in\{1,3,6,12\}$ with the same total workload.
The Passive HBF baseline runs on the GPU, fetching weights and state from HBF on HBM misses.
QKV projection and routing remain on the GPU.

\noindent\textbf{Metrics and energy.}
Mean E2E latency averages workflow arrival-to-final-agent completion times, including queueing.
Throughput is total output tokens across all agents divided by elapsed time from the first workflow's arrival to the last completion.
GPU energy uses execution times and NVML board power (including HBM); host, HBF, and transfer energy are modeled~\cite{hbfchar2026}.
Host and HBF background power apply throughout execution; HBF logic power applies during HBF computation.

\subsection{Comparison with Prior Work}
\label{sec:eval_prior}
\begin{figure}[t]
\SynapseFloatColumn{R}
\centering
\includegraphics[width=\columnwidth]{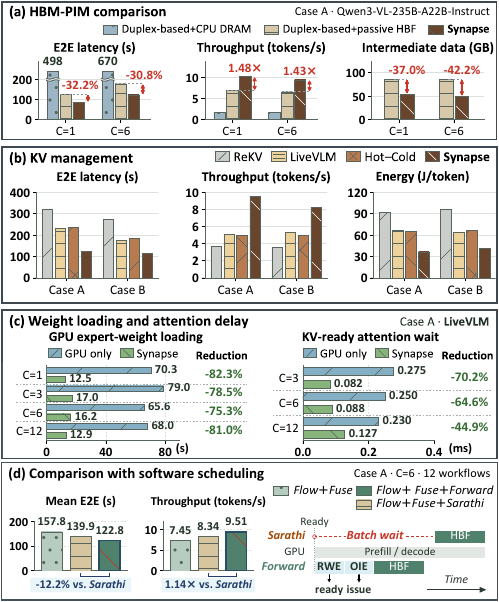}
\caption{Synapse versus prior work: (a) HBM-PIM; (b) KV management; (c) weight loading and attention delay; (d) software  scheduling.}
\Description{Synapse versus prior work: (a) HBM-PIM; (b) KV management; (c) weight loading and attention delay; (d) software  scheduling.}
\label{fig:eval_prior}
\end{figure}

\para{HBM-PIM.}
Fig.~\ref{fig:eval_prior}(a) compares Synapse with Duplex-based CPU and passive HBF configurations~\cite{yun2024duplex} for Case A at $C=1,6$.
All use 80\,GB HBM and 2,048 BF16 MACs at 400\,MHz (1.64\,TFLOPS).
The CPU baseline holds all weights in 512\,GB DRAM; both Duplex variants load expert weights into HBM as needed and send intermediate results to the GPU to be summed.
Against Duplex with passive HBF, Synapse reduces E2E by 30.8--32.2\%, delivers 1.43--1.48$\times$ throughput, and sends 37.0--42.2\% less intermediate data to the GPU.
\textit{Flow}/\textit{Fuse} reduce GPU transfers, while \textit{Forward} reduces command waits, accelerating workflows beyond capacity expansion alone.

\para{KV management.}
At $C=6$, both workflows use 256 frames per video, with retrieval and transfers included.
ReKV retrieves 64 blocks~\cite{rekv}; LiveVLM retains 4,896 tokens after compression and retrieval~\cite{livevlm}; Hot--Cold uses LRU placement with write-on-eviction~\cite{hotcold2026}.
Synapse retains full video KV while reducing E2E by 35.7--61.8\% (Fig.~\ref{fig:eval_prior}(b)).
With ReKV/LiveVLM selections and expert routes fixed, Synapse still reduces E2E by 38.2--40.2\%, separating execution gains from KV selection.

\para{Weight loading and attention delay.}
With LiveVLM selections and expert routes fixed, \textit{Fuse} executes experts in HBF to avoid GPU weight transfers, reducing cumulative loading time by 75.3--82.3\% at $C=1,3,6,12$ (Fig.~\ref{fig:eval_prior}(c)).
\textit{Flow} executes ready attention on HBF, while \textit{Fuse} reduces competing GPU work.
Across 7,090 identical KV-ready tasks, Synapse reduces mean ready-to-start delay by 44.9--70.2\% at $C=3,6,12$.
At $C=12$, GPU weight loading remains 81.0\% lower, showing that expert execution in HBF retains its benefit despite the smaller reduction in attention wait.

\para{Software scheduling.}
Sarathi-Serve combines chunked prefills with stall-free batching~\cite{sarathi_osdi2024}.
With \textit{Flow}/\textit{Fuse} enabled, we replay its policy at $C=6$ with modeled HBF, the same GPU/HBF assignment rules, and $4\,\mu$s command service (Fig.~\ref{fig:eval_prior}(d)).
Sarathi reduces mean E2E from 157.8 to 139.9\,s.
Compared with Sarathi, \textit{Forward} achieves 122.8\,s E2E and
9.51 tokens/s throughput, improving both by 12.2\% and 14.0\%, respectively.
RWE marks successors ready after HBM write acknowledgment once all required inputs are available; OIE issues ready HBF work independently of the GPU, addressing the dependency gaps in Fig.~\ref{fig:synapse_motivation}(c).
With command service time held constant, \textit{Forward} allows ready HBF work to start while the GPU processes another agent's prefill, shortening the gap between data readiness and execution.

\subsection{Ablation Studies and Resource Efficiency}
\label{sec:eval_ablation}
\label{sec:eval_resources}
\begin{figure}[t]
\SynapseFloatColumn{L}
\centering
\includegraphics[width=\columnwidth]{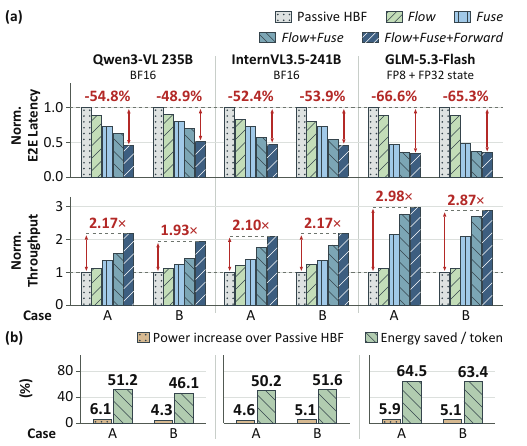}
\caption{Ablation: (a) latency and throughput normalized to Passive HBF; (b) mean power increase and energy savings per token for the full configuration.}
\Description{Ablation: (a) latency and throughput normalized to Passive HBF; (b) mean power increase and energy savings per token for the full configuration.}
\label{fig:eval_ablation}
\label{fig:eval_resources}
\end{figure}

\para{Latency and throughput.}
The five configurations in Fig.~\ref{fig:eval_ablation}(a) test complementary benefits from attention offload, expert offload, and command handling with fixed 64-frame workloads, 2,048 MACs, and buffers.
This ablation also supports HBF execution of HBM-cached experts; for GLM, \textit{Flow} additionally executes Manifold-Constrained Hyper-Connections (mHC) and its associated normalization on the VFE.
\textit{Flow}+\textit{Fuse} lowers E2E latency more than either alone in all six model--workflow cases, showing that the two features address complementary GPU bottlenecks (Fig.~\ref{fig:synapse_flow_fuse}).
The full configuration reduces E2E latency by 48.92--66.60\% and increases throughput to $1.93$--$2.98\times$ that of Passive HBF.

\para{Issue and completion.}
With \textit{Flow}/\textit{Fuse} enabled, the comparison fixes the GPU/HBF assignment rule and assumed $4\,\mu$s command service time.
\textit{Forward} further reduces E2E latency by up to 28.36\%: OIE issues ready HBF work without waiting for GPU command processing, while RWE marks successors ready as soon as their inputs are written to HBM (Fig.~\ref{fig:synapse_execution}).

\para{Compute scaling.}
The MAC sweep tests whether gains require a larger HBF engine, keeping vector lanes, buffers, and link bandwidth fixed.
With 2,048 MACs, at least 88.7\% of the 8,192-MAC latency reduction over Passive HBF is retained. For GLM, E2E latency changes by at most 0.32\%.
Thus, most of the benefit is achieved without quadrupling MAC resources.

\para{Energy per output token.}
On identical traces, full Synapse reduces modeled energy per output token by 46.13--64.48\% versus Passive HBF (Fig.~\ref{fig:eval_resources}(b)).
Although modeled mean power increases by 4.31--6.08\%, Synapse finishes sooner and therefore consumes less total energy.

\subsection{Flash Endurance with \textit{Fortify}}
\label{sec:eval_endurance}
\begingroup
\setlength{\intextsep}{4pt plus 1pt minus 1pt}
\begin{figure}[H]
\centering
\includegraphics[width=\columnwidth]{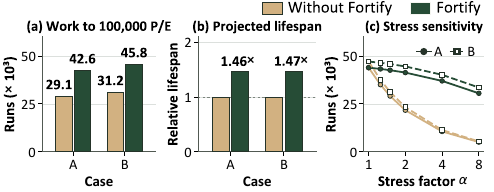}
\caption{\emph{Fortify} with a 100,000-P/E budget: (a) work completed, (b) projected relative lifespan at equal workload rates, and (c) sensitivity to stress factor $\alpha$.}
\Description{\emph{Fortify} with a 100,000-P/E budget: (a) work completed, (b) projected relative lifespan at equal workload rates, and (c) sensitivity to stress factor $\alpha$.}
\label{fig:eval_fortify}
\end{figure}
\endgroup

Refresh mitigates read disturb by rewriting valid pages, consuming NAND P/E cycles.
The HBF endurance model~\cite[Eq.~(8)]{hbfchar2026} projects write-limited lifespan from NAND-access traces as
\begin{equation}
L_{\mathrm{write}}(\alpha)=\frac{P C_F}{W_F(\alpha)/T},
\label{eq:eval_lifespan}
\end{equation}
where $P$ is the P/E budget, $C_F$ the wear-leveling capacity, and $W_F(\alpha)$ the physical writes over $T$, including stress-dependent refresh writes.
Comparing block-level control with \textit{Fortify}'s wordline tracking tests whether finer accounting avoids premature refresh~\cite{fortify_straw}.
Both replay identical Qwen3-VL 235B traces with matched safety margins, banks, spares, and refresh accounting; unused capacity is excluded.
The read-count and effective-stress limits are both 100,000.
The stress factor $\alpha$ denotes disturbance from an adjacent-wordline read relative to a nonadjacent read; both policies use $\alpha\in\{1,1.25,1.5,2,4,8\}$.
Replay ends when the first weight block reaches 100,000 P/E cycles; each run comprises 12 workflows.
At $\alpha=1.5$, \textit{Fortify} completes 46.04--46.77\% more runs before reaching the P/E limit, giving the same projected lifetime gain at equal workload rates (Fig.~\ref{fig:eval_fortify}(a--b)).
Higher $\alpha$ shortens both lifetimes, but \textit{Fortify} limits this loss by tracking each wordline's remaining stress margin and avoiding redundant refreshes of shared weights (Fig.~\ref{fig:eval_fortify}(c)).

\subsection{Area Estimation}
\label{sec:eval_area}
\begin{table}[tbp]
\SynapseFloatColumn{R}
\makeatletter
\long\def\@makecaption#1#2{%
  \footnotesize\bgroup\par\centering
  \@IEEEtabletopskipstrut
  {\normalfont\footnotesize #1:\nobreakspace\scshape #2}\par
  \addvspace{0.5\baselineskip}\egroup
  \@IEEEtablecaptionsepspace
}
\makeatother
\centering
\caption{Synapse Component Area Estimation}
\label{tab:eval_area}
\small
\setlength{\tabcolsep}{3pt}
\begin{tabularx}{\columnwidth}{@{}Xr@{}}
\toprule
\textbf{Structure (five HBM--HBF pairs in total)}
& \textbf{Area (mm$^2$)} \\
\midrule
\textbf{HBM base-die additions} & \textbf{0.120} \\
Operand Issue Engine (OIE; 5 units, 20 KiB) & 0.047 \\
Result Writeback Engine (RWE; 5 units, 20 KiB) & 0.033 \\
Physical implementation allowance & 0.040 \\
\midrule
\textbf{HBF base-die additions} & \textbf{72.801} \\
Shared MAC Engine (SME; 2,048 MACs) & 0.821 \\
Vector Function Engine (VFE; 16 lanes) & 0.643 \\
Shared working SRAM (6.75 MiB) & 4.126 \\
Attention score SRAM (6 MiB) & 3.060 \\
Other SRAM and register files (3.25 MiB) & 1.944 \\
HBF control and metadata (128 MiB) & 56.036 \\
Physical implementation allowance & 6.170 \\
\midrule
\textbf{Total area budget (7 nm)} & \textbf{72.92} \\
\midrule
HBF budget / five HBF stack footprints & $\approx 8.29\%$ \\
\bottomrule
\end{tabularx}
\end{table}

Table~\ref{tab:eval_area} estimates added area across five HBM--HBF pairs.
Shared \emph{Flow}/\emph{Fuse} hardware reuses NAND page buffers; working SRAM includes 2 MiB of FP32 partial sums.
CACTI SRAM estimates at 22\,nm are projected to 7\,nm using FLINT's ratio ($3.9/14.3$)~\cite{synapse_area_cacti,fortify_flint}.
MAC/control references use ASAP7; VFE uses an FPnew-based allowance~\cite{synapse_area_asap7,synapse_area_fpnew}.
\emph{Forward} reserves 64 task slots per pair; RWE tracks up to four issued HBF tasks until writeback completes.
For the evaluated Qwen layout, \emph{Fortify} histories and metadata round up to 64 KiB per bank, provisioned independently across 2,048 active banks (128 MiB).
SRAM ECC, 2\,$\mu$m placement margins around SRAM macros, 10\% cell-area buffering, and 70\% standard-cell utilization are included.
The estimate is 72.92\,mm$^2$: 0.120\,mm$^2$ on HBM
and 72.801\,mm$^2$ on HBF.
The latter is 8.29\% of five nominal HBF footprints (878\,mm$^2$)~\cite{ocp}.
\section{Conclusion}
\label{sec:conclusion}
Synapse overlaps agent execution across the GPU and HBF.
Its shared HBF engine runs attention using KV in HBM (\textit{Flow}) and experts using weights in NAND (\textit{Fuse}).
\textit{Forward} issues ready HBF work from the HBM controller and marks dependent tasks ready when results reach HBM.
\textit{Fortify} counts physical NAND reads across agents to avoid unnecessary refreshes of shared weights.
H100-profiled simulations show 48.92--66.60\% lower E2E latency, up to $2.98\times$ throughput, and up to 64.48\% lower modeled energy per output token than Passive HBF.
Wear simulations project 46--47\% longer HBF lifespan at the same workload rate.

\clearpage
\bibliographystyle{IEEEtran}
\bibliography{refs}

@misc{gemini_agentic_video2026,
  author = {Rohan Doshi and Mario Lu{\v{c}}i{\'c}},
  title = {{Introducing agentic video understanding with Gemini}},
  howpublished = {Google},
  year = {2026},
  month = sep,
  url = {https://blog.google/innovation-and-ai/models-and-research/gemini-models/introducing-agentic-video-in-gemini/}
}

@article{qwen25vl_report,
  author = {{Qwen Team}},
  title = {{Qwen2.5-VL Technical Report}},
  journal = {arXiv preprint arXiv:2502.13923},
  year = {2025},
  month = feb,
  eprint = {2502.13923},
  archivePrefix = {arXiv},
  primaryClass = {cs.CV},
  doi = {10.48550/arXiv.2502.13923},
  url = {https://arxiv.org/abs/2502.13923}
}

@article{qwen3vl_report,
  author = {{Qwen Team}},
  title = {{Qwen3-VL Technical Report}},
  journal = {arXiv preprint arXiv:2511.21631},
  year = {2025},
  month = nov,
  eprint = {2511.21631},
  archivePrefix = {arXiv},
  primaryClass = {cs.CV},
  doi = {10.48550/arXiv.2511.21631},
  url = {https://arxiv.org/abs/2511.21631}
}

@inproceedings{a4vl2026,
  author = {Yichang Xu and Gaowen Liu and Ramana Rao Kompella and Tiansheng Huang and Sihao Hu and Fatih Ilhan and Selim Furkan Tekin and Zachary Yahn and Ling Liu},
  title = {A Multi-Agent Perception-Action Alliance for Efficient Long Video Reasoning},
  booktitle = {Proceedings of the IEEE/CVF Conference on Computer Vision and Pattern Recognition (CVPR)},
  year = {2026},
  url = {https://arxiv.org/abs/2603.14052}
}

@misc{ocp,
  author = {Chinnakrishnan Ballapuram and Dongsop Lee},
  title = {{High Bandwidth Flash (HBF): High-Level Base Die Specification, Version 0.7.0}},
  year = {2026},
  month = aug,
  howpublished = {Open Compute Project},
  url = {https://www.opencompute.org/documents/ocp-hbf-architecture-specification-v0-7-0-final-pdf}
}

@InProceedings{ego4d_cvpr2022,
  author = {Grauman, Kristen and Westbury, Andrew and Byrne, Eugene and Chavis, Zachary and Furnari, Antonino and Girdhar, Rohit and Hamburger, Jackson and Jiang, Hao and Liu, Miao and Liu, Xingyu and Martin, Miguel and Nagarajan, Tushar and Radosavovic, Ilija and Ramakrishnan, Santhosh Kumar and Ryan, Fiona and Sharma, Jayant and Wray, Michael and Xu, Mengmeng and Xu, Eric Zhongcong and Zhao, Chen and Bansal, Siddhant and Batra, Dhruv and Cartillier, Vincent and Crane, Sean and Do, Tien and Doulaty, Morrie and Erapalli, Akshay and Feichtenhofer, Christoph and Fragomeni, Adriano and Fu, Qichen and Gebreselasie, Abrham and Gonz\'alez, Cristina and Hillis, James and Huang, Xuhua and Huang, Yifei and Jia, Wenqi and Khoo, Weslie and Kol\'a\v{r}, J\'achym and Kottur, Satwik and Kumar, Anurag and Landini, Federico and Li, Chao and Li, Yanghao and Li, Zhenqiang and Mangalam, Karttikeya and Modhugu, Raghava and Munro, Jonathan and Murrell, Tullie and Nishiyasu, Takumi and Price, Will and Ruiz, Paola and Ramazanova, Merey and Sari, Leda and Somasundaram, Kiran and Southerland, Audrey and Sugano, Yusuke and Tao, Ruijie and Vo, Minh and Wang, Yuchen and Wu, Xindi and Yagi, Takuma and Zhao, Ziwei and Zhu, Yunyi and Arbel\'aez, Pablo and Crandall, David and Damen, Dima and Farinella, Giovanni Maria and Fuegen, Christian and Ghanem, Bernard and Ithapu, Vamsi Krishna and Jawahar, C. V. and Joo, Hanbyul and Kitani, Kris and Li, Haizhou and Newcombe, Richard and Oliva, Aude and Park, Hyun Soo and Rehg, James M. and Sato, Yoichi and Shi, Jianbo and Shou, Mike Zheng and Torralba, Antonio and Torresani, Lorenzo and Yan, Mingfei and Malik, Jitendra},
  title = {{Ego4D}: Around the World in 3,000 Hours of Egocentric Video},
  booktitle = {Proceedings of the IEEE/CVF Conference on Computer Vision and Pattern Recognition (CVPR)},
  month = {June},
  year = {2022},
  pages = {18995--19012},
  url = {https://openaccess.thecvf.com/content/CVPR2022/html/Grauman_Ego4D_Around_the_World_in_3000_Hours_of_Egocentric_Video_CVPR_2022_paper.html}
}

@article{hotcold2026,
  author = {Jongjin Baek and Won Ji and Seungjae Yoo and Joo-Young Kim},
  title = {{Hot--Cold Tiering of HBM and High Bandwidth Flash for Agentic LLM Serving}},
  journal = {IEEE Computer Architecture Letters},
  volume = {25},
  number = {2},
  pages = {355--358},
  year = {2026},
  doi = {10.1109/LCA.2026.3729099},
  url = {https://doi.org/10.1109/LCA.2026.3729099}
}

@inproceedings{yun2024duplex,
  author = {Sungmin Yun and Kwanhee Kyung and Juhwan Cho and Jaewan Choi and Jongmin Kim and Byeongho Kim and Sukhan Lee and Kyomin Sohn and Jung Ho Ahn},
  title = {{Duplex: A Device for Large Language Models with Mixture of Experts, Grouped Query Attention, and Continuous Batching}},
  booktitle = {2024 57th IEEE/ACM International Symposium on Microarchitecture (MICRO)},
  year = {2024},
  pages = {1429--1443},
  doi = {10.1109/MICRO61859.2024.00105},
  url = {https://doi.org/10.1109/MICRO61859.2024.00105}
}

@inproceedings{rekv,
  author = {Shangzhe Di and Zhelun Yu and Guanghao Zhang and Haoyuan Li and Tao Zhong and Hao Cheng and Bolin Li and Wanggui He and Fangxun Shu and Hao Jiang},
  title = {{Streaming Video Question-Answering with In-context Video KV-Cache Retrieval}},
  booktitle = {International Conference on Learning Representations (ICLR)},
  year = {2025},
  url = {https://arxiv.org/abs/2503.00540}
}

@inproceedings{livevlm,
  author = {Zhenyu Ning and Guangda Liu and Qihao Jin and Chengwei Li and Wenchao Ding and Minyi Guo and Jieru Zhao},
  title = {{LiveVLM: Efficient Online Video Understanding via Streaming-Oriented KV Cache and Retrieval}},
  booktitle = {63rd ACM/IEEE Design Automation Conference (DAC)},
  year = {2026},
  month = jul,
  pages = {1--7},
  doi = {10.1145/3770743.3804012},
  url = {https://doi.org/10.1145/3770743.3804012}
}

@inproceedings{pregated,
  author = {Ranggi Hwang and Jianyu Wei and Shijie Cao and Changho Hwang and Xiaohu Tang and Ting Cao and Mao Yang},
  title = {{Pre-gated MoE}: An Algorithm-System Co-Design for Fast and Scalable Mixture-of-Expert Inference},
  booktitle = {2024 ACM/IEEE 51st Annual International Symposium on Computer Architecture (ISCA)},
  year = {2024},
  pages = {1018--1031},
  doi = {10.1109/ISCA59077.2024.00078},
  url = {https://doi.org/10.1109/ISCA59077.2024.00078}
}

@article{fortify_straw,
  author = {Myoungjun Chun and Jaeyong Lee and Inhyuk Choi and Jisung Park and Myungsuk Kim and Jihong Kim},
  title = {{STRAW}: A Stress-Aware {WL}-Based Read Reclaim Technique for High-Density {NAND} Flash-Based {SSDs}},
  journal = {IEEE Computer Architecture Letters},
  volume = {24},
  number = {1},
  pages = {5--8},
  year = {2025},
  doi = {10.1109/LCA.2024.3516205},
  url = {https://doi.org/10.1109/LCA.2024.3516205}
}

@article{fortify_flint,
  author = {Geraldo F. Oliveira and Arash Tavakkol and Xiangyu Zhu and Ahmet Caner Y{\"u}z{\"u}g{\"u}ler and Vamanan Arulchelvan and Lukas Cavigelli and Renzo Andri and Mohammad Sadrosadati and Jia Xinglei and Onur Mutlu and Zhou Ke and Shai Bergman and Ji Zhang},
  title = {{FLINT}: Efficiently Leveraging High Bandwidth Flash for Capacity-Scalable {LLM} Inference Acceleration},
  journal = {arXiv preprint arXiv:2608.25062},
  year = {2026},
  month = aug,
  eprint = {2608.25062},
  archivePrefix = {arXiv},
  primaryClass = {cs.AR},
  doi = {10.48550/arXiv.2608.25062},
  url = {https://arxiv.org/abs/2608.25062}
}

@misc{qwen3vl235_config,
  author = {{Qwen Team}},
  title = {{Qwen3-VL-235B-A22B-Instruct}: model configuration},
  year = {2025},
  url = {https://huggingface.co/Qwen/Qwen3-VL-235B-A22B-Instruct/blob/main/config.json},
  note = {Accessed November 2, 2026}
}

@misc{internvl35_241_config,
  author = {{OpenGVLab}},
  title = {{InternVL3.5-241B-A28B}: model configuration},
  year = {2025},
  howpublished = {Official model repository on Hugging Face},
  url = {https://huggingface.co/OpenGVLab/InternVL3_5-241B-A28B/blob/main/config.json},
  note = {Accessed November 7, 2026}
}

@misc{glm53_flash_config,
  author = {{Z.ai}},
  title = {{GLM-5.3-Flash}: model configuration},
  year = {2026},
  howpublished = {Official model repository on Hugging Face},
  url = {https://huggingface.co/zai-org/GLM-5.3-Flash/blob/main/config.json},
  note = {Accessed November 7, 2026}
}

@article{hbfchar2026,
  author = {Zack Yu and Chloe Wong and Coleman Hooper and Minjae Lee and Wonjun Kang and Youngjin Cho and Michael W. Mahoney and Yakun Sophia Shao and Kurt Keutzer and Amir Gholami},
  title = {{Characterizing High Bandwidth Flash for LLM Serving}},
  journal = {arXiv preprint arXiv:2609.39131},
  year = {2026},
  eprint = {2609.39131},
  archivePrefix = {arXiv},
  primaryClass = {cs.LG},
  doi = {10.48550/arXiv.2609.39131},
  url = {https://arxiv.org/abs/2609.39131}
}

@misc{synapse_area_cacti,
  author = {{Hewlett Packard}},
  title = {{CACTI} 7: An Integrated Cache and Memory Access Time, Cycle Time, Area, Leakage, and Dynamic Power Model},
  howpublished = {Open-source modeling tool},
  url = {https://github.com/HewlettPackard/cacti},
  note = {Commit 1ffd8dfb10303d306ecd8d215320aea07651e878, accessed November 9, 2026}
}

@article{synapse_area_fpnew,
  author = {Stefan Mach and Fabian Schuiki and Florian Zaruba and Luca Benini},
  title = {{FPnew}: An Open-Source Multi-Format Floating-Point Unit Architecture for Energy-Proportional Transprecision Computing},
  journal = {arXiv preprint arXiv:2007.01530},
  year = {2020},
  url = {https://arxiv.org/abs/2007.01530}
}

@inproceedings{synapse_area_asap7,
  author = {V. Vashishtha and M. Vangala and L. T. Clark},
  title = {{ASAP7} Predictive Design Kit Development and Cell Design Technology Co-Optimization: Invited Paper},
  booktitle = {IEEE/ACM International Conference on Computer-Aided Design (ICCAD)},
  pages = {992--998},
  year = {2017},
  doi = {10.1109/ICCAD.2017.8203889},
  url = {https://doi.org/10.1109/ICCAD.2017.8203889}
}

@misc{synapse_h100_arch,
  author = {{NVIDIA}},
  title = {{NVIDIA H100 Tensor Core GPU Architecture}},
  year = {2022},
  howpublished = {Architecture white paper, version 1.04},
  url = {https://dam-cdn.nvd.orangelogic.com/AssetLink/705n6ur546g0uk43w0117r17n8042d73.pdf},
  note = {Accessed November 16, 2026}
}

@inproceedings{longvideobench,
  author = {Haoning Wu and Dongxu Li and Bei Chen and Junnan Li},
  title = {{LongVideoBench}: A Benchmark for Long-context Interleaved Video-Language Understanding},
  booktitle = {Advances in Neural Information Processing Systems},
  volume = {37},
  pages = {28828--28857},
  year = {2024},
  doi = {10.52202/079017-0907},
  url = {https://proceedings.neurips.cc/paper_files/paper/2024/file/329ad516cf7a6ac306f29882e9c77558-Paper-Datasets_and_Benchmarks_Track.pdf}
}

@inproceedings{sarathi_osdi2024,
  author = {Amey Agrawal and Nitin Kedia and Ashish Panwar and Jayashree Mohan and Nipun Kwatra and Bhargav Gulavani and Alexey Tumanov and Ramachandran Ramjee},
  title = {Taming {Throughput-Latency} Tradeoff in {LLM} Inference with {Sarathi-Serve}},
  booktitle = {18th USENIX Symposium on Operating Systems Design and Implementation (OSDI 24)},
  year = {2024},
  isbn = {978-1-939133-40-3},
  address = {Santa Clara, CA},
  pages = {117--134},
  publisher = {USENIX Association},
  month = jul,
  url = {https://www.usenix.org/conference/osdi24/presentation/agrawal}
}

\end{document}